\documentclass[a4paper,12pt]{article} 
\usepackage{amsmath,amsthm,amssymb}
\newcommand{\bm}[1]{\mbox{\boldmath$#1$}}
\newcommand{\be}{\begin{equation}}
\newcommand{\ee}{\end{equation}}
\newcommand{\bea}{\begin{eqnarray}}
\newcommand{\eea}{\end{eqnarray}}
\newcommand{\non}{\nonumber}

\newtheorem{th1}{Theorem}
\newtheorem{lem}{Lemma}
\newtheorem{conj}{Conjecture}
\newtheorem{prop}{Proposition}

\newcommand{\ra}{\rangle}

\newcommand{\al}{\alpha}

\begin{document}

\title{The six-vertex model at roots of unity
and some highest weight representations of 
the $sl_2$ loop algebra}

\author{Tetsuo Deguchi
\footnote{e-mail deguchi@phys.ocha.ac.jp}}

\date{}

\maketitle

\begin{center}  
Department of Physics, Ochanomizu University, \\
2-1-1 Ohtsuka, Bunkyo-ku,Tokyo 112-8610, Japan
\end{center} 
\abstract{
We discuss irreducible highest weight representations 
of the $sl_2$ loop algebra and reducible indecomposable ones 
in association with the $sl_2$ loop algebra symmetry 
of the six-vertex model at roots of unity. 
We formulate an elementary proof that every highest weight representation 
with distinct evaluation parameters is irreducible. 
We present a general criteria for a highest weight representation 
to be irreducble. We also give an example of a  
reducible indecomposable highest weight representation 
and discuss its dimensionality. }

\section{Introduction}

Roots of unity representations of the quantum groups have 
 many subtle and interesting properties, 
 as was first discussed by Roche and Arnaudon \cite{Arnaudon,DeConcini}.     
 They should have significant implications in integrable systems.   
In fact, the roots of unity representation theory has been utilized to 
generalize the R matrices of the chiral Potts model \cite{Kyoto}. 
There have been several approaches to  connect 
 roots of unity representations to the six-vertex model 
 and the XXZ spin chain \cite{Rittenberg,Pasquier}. 
 These studies could be fundamental and should  
 contain useful suggestions for future researches,  
 since the six-vertex model is one of the most important exactly solvable 
 models in statistical mechanics. 
Recently, it was shown that  
 the six-vertex model at roots of unity 
 has many spectral degeneracies which are associated 
 with the $sl_2$ loop algebra \cite{DFM}. Furthermore, it has been 
 found that the $sl_2$ loop algebra symmetry 
is closely related to roots of unity representations that are  
 discussed by Lusztig \cite{Modular,Chari-P2}.

Let us introduce the $sl_2$ loop algebra symmetry of the 
six-vertex model at roots of unity. 
It was explicitly shown that when $q$ is a root of unity  
the transfer matrix of the six-vertex model commutes 
 with the generators of the $sl_2$ loop algebra \cite{DFM}.   
Let $q_0$ be a primitive root of unity satisfying 
$q_0^{2N}=1$ for an integer $N$. 
We introduce operators $S^{\pm(N)}$ as follows 
\begin{eqnarray}
S^{\pm(N)} &=&  
\sum_{1 \le j_1 < \cdots < j_N \le L}
q_0^{{\frac N 2 } \sigma^Z} \otimes \cdots \otimes q_0^{{N \over 2} \sigma^Z}
\otimes \sigma_{j_1}^{\pm} \otimes
q_0^{{(N-2) \over 2} \sigma^Z} \otimes  \cdots \otimes q_0^{{(N-2) \over 2}
\sigma^Z}
\nonumber \\
 & & \otimes \sigma_{j_2}^{\pm} \otimes q_0^{{(N-4) \over 2} \sigma^Z} \otimes
\cdots
\otimes \sigma^{\pm}_{j_N} \otimes q_0^{-{N \over 2} \sigma^Z} \otimes \cdots
\otimes q_0^{-{N \over 2} \sigma^Z}  \, . 
\label{sn}
\end{eqnarray}
The operators $S^{\pm(N)}$ are derived from the $N$th power 
of the generators $S^{\pm}$ of the quantum group $U_q(sl_2)$ or 
$U_q(\hat{sl_2})$. 
We also define  $T^{(\pm)}$ by the complex conjugates of $S^{\pm(N)}$,  
i.e. $T^{\pm (N)} = \left( S^{\pm (N)} \right)^{*}$. 
The operators,  $S^{\pm(N)}$ and $T^{\pm (N)}$, 
generate the $sl_2$ loop algebra, $U(L(sl_2))$, in the sector 
\begin{equation}
 S^{Z} \equiv 0 \quad ({\rm mod} \, N) \, . 
\label{sct0}
\end{equation}
 Here the value of the total spin 
$S^Z$ is given by an integral multiple of $N$. 
In the sector (\ref{sct0}),  the operators 
$S^{\pm (N)}$ and $T^{\pm (N)}$ (anti-)commute with 
the  transfer matrix of the six-vertex model $\tau_{6V}(v)$.

The loop algebra symmetry is also in common with 
the XXZ spin chain, which is one of the most significant 
quantum integrable systems. We note that  
the logarithmic derivative 
of the transfer matrix of the six-vertex model 
gives the Hamiltonian of the XXZ spin chain  
under the periodic boundary conditions:  
\begin{equation} 
H_{XXZ} =  {\frac 1 2} \sum_{j=1}^{L} \left(\sigma_j^X \sigma_{j+1}^X +
 \sigma_j^Y \sigma_{j+1}^Y + \Delta \sigma_j^Z \sigma_{j+1}^Z  \right) \, . 
\label{hxxz}
\end{equation}
Here the XXZ anisotropic coupling $\Delta$ 
is related to the $q$ parameter by  $\Delta= (q+q^{-1})/2$. 
In the sector (\ref{sct0}), the operators 
$S^{\pm (N)}$ and $T^{\pm (N)}$ 
commute with the Hamiltonian of the XXZ spin chain: \cite{DFM}  
\begin{equation}
{[}S^{\pm(N)},H_{XXZ} {]}={[}T^{\pm(N)},H_{XXZ} {]}=0 \, . 
\label{sthcomm}
\end{equation} 

Let us now discuss the spectral degeneracy of the $sl_2$ loop algebra. 
In the sector (\ref{sct0}) every Bethe ansatz eigenvector $|B \ra$ 
may have the following degenerate eigenvectors 
$$
S^{- (N)} | B \rangle \, , \quad T^{- (N)} | B \rangle \, ,  
\quad (S^{-(N)})^2 | B \rangle \, ,  \quad  
T^{- (N)} S^{+(N)} T^{- (N)} | B \rangle \, , \cdots \, . 
$$
However, it is nontrivial how many of them are linearly independent.  
The number should explain the degree of the spectral degeneracy. 
Thus, we want to know the dimensions of representations 
generated by Bethe ansatz eigenvectors. 

Fabricius and McCoy has conjectured that Bethe ansatz eigenvectors 
should be highest weight vectors with respect to the 
$sl_2$ loop algebra symmetry, and also that they have 
Drinfeld polynomials. 
In fact, it has been shown in the sector (\ref{sct0}) that 
a regular Bethe ansatz eigenvector is highest weight \cite{HWT}. 
Furthermore, it has been shown that
 if the evaluation parameters of highest weight vector $\Omega$ 
 are distinct, the highest weight representation generated by $\Omega$ 
is irreducible \cite{Chari-P3}. Here we note that 
highest weight vectors and evaluation parameters are defined in \S 3.1.  
Thus, in the sector (\ref{sct0}), 
if the evaluation parameters of a regular Bethe state 
are distinct, the highest weight representation is irrducible and 
it has the Drinfeld polynomial by which we can determine 
the dimensions. However, if the evaluation parameters are not distinct, 
it is not trivial whether the highest weight representation 
is irrducible or not. We emphasize that 
a finite-dimensional highest weight representation 
is not necessarily irreducible.

In the paper, we discuss a different proof of the theorem 
that every finite-dimensional highest weight representation 
with distinct evaluation parameters is irreducible. 
We introduce an elementary computational scheme, and  
construct explicitly the basis of such a representation 
with distinct evaluation parameters. 
Furthermore, we show  a general criteria 
for a finite-dimensional highest weight representation to be irreducible. 
We then  discuss briefly how to apply them 
to the spectral degeneracy of the six-vertex model at roots of unity. 
Finally, we give an  example of a 
reducible highest weight representation.

\section{Loop algebra generators with parameters }

We consider the classical analogue of the Drinfeld 
realization of the quantum $sl_2$ loop algebra, 
$U_q(L(sl_2))$. 
The classical analogues of the Drinfeld generators, ${x}_k^{\pm}$ and 
${h}_k$ ($k \in {\bf Z}$), 
 satisfy the defining relations: 
\begin{equation} 
[{h}_j, {x}_{k}^{\pm} ] = \pm 2 {x}_{j+k}^{\pm} \, , \quad 
[{x}_j^{+}, {x}_k^{-} ] = {h}_{j+k} \, , 
\quad {\rm for} \, j, k 
\in {\bf Z} \, . 
\label{CDR}
\end{equation}
Here $[{ h}_j, {h}_{k} ]=0$ and 
$[{x}_j^{\pm}, {x}_k^{\pm}] =0$ 
for $j, k \in {\bf Z}$.  

Let $A$ be a set of parameters such as 
$\{\alpha_1, \alpha_2, \ldots, \alpha_m \}$. 
We define generators with $m$ parameters $x_m^{\pm}(A)$ and 
$h_m(A)$ as follows 
\begin{eqnarray} 
x_{m}^{\pm}(A) & = & 
\sum_{k=0}^{m} (-1)^k x_{m-k}^{\pm} 
\sum_{\{ i_1, \ldots, i_k \} \subset \{1, \ldots, m \}}  
\alpha_{i_1} \alpha_{i_2} \cdots \alpha_{i_k} \, , 
\non \\
h_{m}(A) & = & 
\sum_{k=0}^{m} (-1)^k h_{m-k}^{\pm} 
\sum_{\{ i_1, \ldots, i_k \} \subset \{1, \ldots, m \}}  
\alpha_{i_1} \alpha_{i_2} \cdots \alpha_{i_k} \, . 
\end{eqnarray}
In terms of generators with parameters 
we generalize the defining relations of the $sl_2$ loop algebra. 
Let $A$ and $B$ are arbitrary sets of $m$ and $n$ parameters, 
respectively.  The operators with parameters satisfy the following: 
\begin{equation} 
[x_m^{+}(A), x_n^{-}(B)] = h_{m+n}(A \cup B) \, , \quad 
[h_{m}(A), x^{\pm}_n(B)] = \pm 2  x_{m+n}^{\pm}(A \cup B) \, . 
\label{eq:dfr-AB}
\end{equation}
They generalize the one-parameter operators introduced in Ref. \cite{HWT}. 
By using  the relations (\ref{eq:dfr-AB}), 
it is straightforward to show the following useful relations: 
\begin{eqnarray} 
 [ x_{\ell}^{+}(A), (x_m^{-}(B))^{(n)} ] & = & (x_m^{-}(B) )^{(n-1)} 
 h_{\ell+m}(A \cup B) - x_{\ell+2m}^{-}(A \cup B \cup B) 
 (x_{m}^{-}(B))^{(n-2)} \, , \non \\ 
 {[} h_{\ell}(A), (x_m^{\pm}(B))^{(n)} {]} & = & \pm 2 (x_m^{\pm}(B))^{(n-1)} 
 x_{\ell+m}^{\pm}(A \cup B) \, . \label{eq:AB}
\end{eqnarray}
Here the symbol $(X)^{(n)}$ denotes the $n$th power of operator $X$ 
divided by the $n$ factorial, i.e. $(X)^{(n)} = X/n!$ .

Let the symbol ${\bm \alpha}$ denote 
a set of $m$ parameters, $\al_j$ for $j=1, 2, \ldots, m$.  
We denote by $A_{j}$ the set of all the parameters  
except for $\al_j$, i.e.  
$A_{j}= {\bm \alpha} \setminus \{\al_j  \}= 
\{\al_1, \ldots, \al_{j-1}, \al_{j+1}, \ldots, \al_m \}$.   
We introduce the following symbol: 
\begin{equation} 
\rho_j({\bm \alpha}) = x_{m-1}^{-}(A_j) \, \quad 
{\rm for } \quad j = 1, 2, \ldots, m . 
\end{equation} 
The generators $x_{j}^{-}$ for $j=0, 1, \ldots, m-1$,  
are expressed as linear combinations of 
$\rho_j({\bm \alpha})$. 
Let us introduce $\al_{kj}=\al_k - \al_j$.  
We show the following: 
\be 
\sum_{j=1}^{n} {\frac {\rho_j({\bm \alpha})} 
{\prod_{k=1; k \ne j}^{m} \al_{kj}} } 
= x_{m-n}^{-}(\{ \al_{n+1}, \ldots, \al_{m} \})  \quad (1 \le n \le m)  \, . 
\ee   
It thus follows inductively on $n$ that $x_{k}^{-}$ ($0 \le k \le m-1$) 
are expressed in terms of linear combinations of  $\rho_j({\bm \alpha})$ 
with $1 \le j \le m$.

\section{Highest weight representations}

\subsection{Evaluation parameters of a highest weight vector}

In a representation of $U(L(sl_2))$, we call a vector $\Omega$ 
{\it a highest weight vector} if  $\Omega$ 
is annihilated by generators ${x}_{k}^{+}$ 
for all integers $k$ and such that 
$\Omega$ is a simultaneous eigenvector of every generator 
of the Cartan subalgebra, ${h}_k$ ($k\in {\bf Z}$)  
\cite{Chari-P1,Chari-P2}: 
\bea 
{x}_k^{+} \Omega &= & 0 \, , \quad {\rm for} \, \, k \in {\bf Z} \, , 
\label{eq:annihilation} \\ 
{h}_{k} \Omega & = & {d}_k^{+} \Omega \, , \quad 
{h}_{-k} \Omega = {d}_{-k}^{-} \Omega \, , 
\quad {\rm for} \, \, k \in {\bf Z}_{\ge 0} \, . 
\label{eq:Cartan}
\eea
The representation generated by a highest weight vector is called 
a highest weight representation. 

Hereafter in the paper, we assume that  
 $\Omega$ is a highest weight vector with highest weight 
$d_k^{\pm}$ and  it generates  a finite-dimensional 
representation. We denote by $V_{\Omega}$ the finite-dimensional 
highest weight representation.

Let us consider the $sl_2$-subalgebra generated by 
$x_{-k}^{+}$, $x_k^{-}$ and $h_0$ for an integer $k$. 
We denote the subalgebra by ${\cal U}_k$. 
Let $r$ be the eigenvalue of ${h}_0$ 
for $\Omega$, i.e. ${h}_0 \Omega= r \Omega$.   
It is easy to show the following: 
\begin{lem} 
The ${\cal U}_k$-subrepresentation 
of $V_{\Omega}$ is given by the $(r+1)$-dimensional irreducible 
representation of $sl_2$, where $r=d^{\pm}_0$.  
\label{prop:sl2sub}
\end{lem}

 For a given integer $n$, 
we define the sector of ${h}_0=r -2n$  
by the vector subspace consisting of vectors $v_n \in V$  
such that ${\bar h}_0 \, v_n= (r - 2n) \, v_n$. 
\begin{lem}
The representation $V_{\Omega}$ is given by 
the direct sum of sectors with respect to eigenvalues of ${h}_0$. 
Furthermore, every vector $v_n$ in the sector of ${h}_0 = r -2n$ 
is expressed as a linear combination of monomial vectors such as 
${x}_{j_1}^{-} \cdots {x}_{j_n}^{-} \, \Omega$ with coefficients 
$C_{j_1, \ldots, j_n}$:
\be 
v_n= \sum_{j_1, \ldots, j_n} C_{j_1, \ldots, j_n} \, 
\prod_{t=1}^{n} {x}_{j_t}^{-} \, \, \Omega
\label{eq:expansion}
\ee
\label{lem:sector} 
\end{lem} 
\begin{proof}    
It is clear from the PBW theorem. 
\end{proof}

Let us now introduce evaluation parameters for $\Omega$. \cite{HWT,loop} 
Using commutation relations of the loop algebra, 
we show that $\Omega$ is a simultaneous eigenvector of  
operators $(x_0^{+})^{(n)}(x_1^{-})^{(n)}$:   
\be 
(x_0^{+})^{(j)}(x_1^{-})^{(j)} \Omega = \lambda_j \Omega \, , \quad 
{\rm for} \quad j=1, 2, \ldots, r \, . 
\label{eq:01}
\ee
In terms of the eigenvalues $\lambda_k$,  
we define a polynomial $P_{\Omega}(u)$ by the following relation: 
\be 
P_{\Omega}(u) =\sum_{k=0}^{r} \lambda_k (-u)^k  \, .   
\label{DrinfeldP}
\ee
Making use of proposition \ref{prop:sl2sub}, we show 
that the roots of  $P_{\Omega}(u)$ 
are nonzero and finite, and the degree of  $P_{\Omega}(u)$ is given by $r$.  
\cite{HWT,loop} 
We note that the author learned the expression of the Drinfeld polynomial 
(\ref{DrinfeldP}) from Jimbo \cite{Jimbo-summer} 
(See also \cite{FM1,Odyssey}). 

Let us factorize $P_{\Omega}(u)$ as follows 
\be 
P_{\Omega}(u) = \prod_{k=1}^{s} (1 - a_k u)^{m_k} \, ,   
\label{eq:factorization}
\ee
where $a_1, a_2, \ldots, a_s$ are distinct, and their  
 multiplicities are given by  $m_1, m_2, \ldots, m_s$, respectively.   
Then, we call  $a_j$ the {\it evaluation parameters} of 
highest weight vector $\Omega$. We denote by ${\bm a}$ the set of 
$s$ parameters, $a_1, a_2, \ldots, a_s$. 
Here we note that $r$ is given by the sum: $r=m_1 + \cdots + m_s$. 
We define parameters ${\hat a}_i$ for $i=1, 2, \ldots, r$, as follows.  
\be 
{\hat a}_i = a_k \quad {\rm if } \, \, m_1+ m_2 + \cdots + m_{k-1} < i \le  
m_1+ \cdots + m_{k-1} + m_{k} \, . 
\label{eq:hat-a}
\ee
Then, the set $\{ {\hat a}_j \, | j =1, 2, \ldots, r \}$ corresponds to the 
set of evaluation parameters $a_j$ with multiplicities $m_j$ for 
 $j=1, 2, \ldots, s$. We denote it by ${\hat A}_{\phi}$.    

It is easy to show the reduction relations 
in the following \cite{Chari-P1,HWT,loop}:  
\begin{lem} 
In the representation $V_{\Omega}$ we have  
\be 
{x}_{\ell + r+1}^{-} \, \Omega = \sum_{j=1}^{r} (-1)^{r-j} 
\lambda_{r+1-j} {x}^{-}_{\ell+ j} \, \Omega \, , \quad {\rm for} \, \, 
\ell \in {\bf Z} \, .  
\label{eq:red-rel-ell}
\ee
Here we recall $r={\bar d}_0^{\pm}$ and $\lambda_j$ denote 
the eigenvalues defined in (\ref{eq:01}). 
\label{lem:red-rel-ell}
\end{lem} 
\begin{proof}
The reduction relation (\ref{eq:red-rel-ell}) is derived from 
the following: 
\bea 
({\bar x}_{-\ell}^{+})^{(n)} ({\bar x}_{\ell+1}^{-})^{(n+1)} 
& = & {\bar x}_{\ell+1}^{-} 
({\bar x}_{-\ell}^{+})^{(n)} ({\bar x}_{\ell+1}^{-})^{(n)} 
+ {\frac 1 2} \, {[} {\bar h}_{1}, ({\bar x}_{-\ell}^{+})^{(n-1)} 
({\bar x}_{\ell+1}^{-})^{(n)} {]} 
\non \\
& & \quad - ({\bar x}_{-\ell}^{+})^{(n-1)} 
({\bar x}_{\ell+1}^{-})^{(n+1)} {\bar x}_{-\ell}^{+} \, , \quad 
 {\rm for} \,\,  \ell \in {\bf Z} \, .  
\label{eq:ind-ell} 
\eea
The relation (\ref{eq:ind-ell}) has been shown for the case of 
$U_q(L(sl(2)))$ \cite{Chari-P1}. 
\end{proof}

The reduction relation (\ref{eq:red-rel-ell}) for $\ell= -1$ 
is expressed as follows 
\be 
x_{r}^{-} ({\hat A}_{\phi}) \Omega = 0 \, . 
\label{eq:red-all}
\ee
Making use of the reduction relation (\ref{eq:red-all}), 
any monomial vector 
$x_{j_1}^{-} x_{j_2}^{-} \cdots x_{j_n}^{-} \Omega$ 
can be expressed as a linear combination of 
$ \rho_{k_1}({\bf a}) \rho_{k_2}({\bf a}) \cdots \rho_{k_n}({\bf a}) \Omega$ 
over some sets of integers $k_1, \ldots, k_n$. 
Here we note the following:  
\begin{lem}
If $x_{n}^{-}(A)\Omega=0$ for some set of parameters, $A$, 
then we have $x_{n}^{-}(A\cup B)\Omega=0$ for any 
set of parameters $B$. 
\label{lem:vanish}
\end{lem}

\subsection{The case of distinct evaluation parameters}

Let us discuss the case where all the evaluation parameters 
$a_j$ have multiplicity 1, i.e. $m_j=1$ for $j=1, \ldots, s$.  
We call it  the case of distinct evaluation parameters. 
Here we note that $s=r$. We therefore have 
\be 
x_{s}^{-} ({\bm a}) \Omega = 0 \, . \label{eq:red-distinct}
\ee
Hereafter, we denote by $a_j^{\otimes m}$ 
the set of parameter $a_j$ with multiplicity $m$, 
i.e. $a_j^{\otimes m}= \{ a_j, a_j, \ldots, a_j \}$.  
For instance, we have  
\be
x_s^{\pm}(a_j^{\otimes s}) = 
\sum_{k=0}^{s} {\frac {s !} {(s-k)! k!}} \, (-a_j)^k x_{s-k}^{\pm} \, .   
\ee
In the case of $s=1$, we write $x_1^{\pm}(a_j^{\otimes 1})$ 
simply as $x_1^{\pm}(a_j)$.

\begin{lem} If all evaluation parameters ${\hat a}_j$ 
are distinct ($m_j=1$ for all $j$), we have 
\be 
\left(\rho_j({\bm a})\right)^2 \, \Omega= 0 \label{eq:vanish} 
\ee
\label{lem:square0}
\end{lem}
\begin{proof} 
 First, we show 
 \be 
 x_0^{+} \,  (\rho_j({\bm a}))^2 \, \Omega= 0 \, . \label{eq:x0}
 \ee
 From eq. (\ref{eq:AB}) we have 
$$ 
x_0^{+} \, (\rho_j({\bm a}))^{(2)} \Omega 
= x_{s-1}^{-}(A_j) h_{s-1}(A_j) \Omega 
- x_{2s-2}^{-} (A_j \cup A_j) \Omega 
$$
In terms of $a_{kj}=a_k - a_j$, we have 
$h_{s-1}(A_j) \Omega = \prod_{k \ne j} a_{kj} \, \Omega$, 
and using eq. (\ref{eq:red-distinct}) and lemma \ref{lem:vanish} 
we have 
$$
x_{2s-2}^{-} (A_j \cup A_j) \Omega = \prod_{k \ne j} a_{kj} \, 
x_{s-1}^{-}(A_j) \Omega \, . 
$$ 
We thus obtain eq. (\ref{eq:x0}).  
Secondly, we apply $(x_0^{+})^{(r-1)}(x_1^{-}(a_j))^{(r-1)}$ to 
$(\rho_j({\bm a}))^2 \Omega$. The product is given by zero 
since it is out of the sectors of $V_{\Omega}$ due to the fact that 
$(r-1)+2 > r$:    
$$ 
(x_0^{+})^{(r-1)}(x_1^{-}(a_j))^{(r-1)} \, 
(\rho_j({\bm a}))^2 \, \Omega= 0 \, . 
$$ 
Here the left-hand-side is given by 
$$
 \rho_j({\bm a})^2 \, (x_0^{+})^{(r-1)}(x_1^{-}(a_j))^{(r-1)} \, \Omega 
 = \prod_{k=1; k \ne j}^{r} a_{kj} \, \times \,
  (\rho_j({\bm a}))^2 \, \Omega 
$$
Since $a_{kj} \ne 0$ for $k \ne j$, 
 we obtain eq. (\ref{eq:vanish}).  
\end{proof}

\begin{lem} 
In the sector $h_0=r- 2n$ of $V_{\Omega}$,  
every vector $v_n$ is expressed as follows:  
\be 
v_n = \sum_{1 \le j_1 \le \cdots \le j_n \le s} 
C_{j_1, \cdots, j_n} \, \prod_{t=1}^{n} \rho_{j_t}({\bm a}) \,  \Omega
\label{eq:vnC}
\ee
If  $v_n$ is zero, 
all the coefficients $C_{j_1, \cdots, j_n}$ are given by zero.  
\label{lem:vnC} 
\end{lem}
\begin{proof} 
By the definition of $\rho_j({\bm a})$, any vector in the sector 
is expressed as a sum over 
$\rho_{j_i}({\bm a})  \cdots \rho_{j_n}({\bm a}) \,  \Omega$.  
We multiply both sides of  eq. (\ref{eq:vnC}) 
with $(x_{\ell}(a_i^{\otimes \ell}))^{(n)}$, and use  
the Vandermonde determinant whose $(i, \ell)$ entry 
is given by $(a_{j_1 i} \cdots a_{j_n i})^{\ell}$ 
to show that if $v_n=0$, all the coefficients $C_{j_1, \cdots, j_n}$ are zero. 
\end{proof} 

We have from lemma \ref{lem:square0} and lemma \ref{lem:vnC} 
the next proposition: 
\begin{prop} If  evaluation parameters ${\hat a}_j$ 
of $\Omega$ are distinct, the set of vectors 
$\prod_{t=1}^{n} \rho_{j_t}({\bm a}) \, \Omega$ for 
$1 \le j_1 \le \cdots \le j_n \le s$ gives a basis 
of the sector $h_0=r- 2n$ of $V_{\Omega}$.    
\label{prop:basis}
\end{prop}

\begin{th1}
Let $\Omega$ be a highest weight vector with distinct evaluation parameters 
$a_1, \ldots, a_r$. The representation generated by $\Omega$ 
is irreducible. 
\label{th:distinct}
\end{th1}
\begin{proof} 
We show that every nonzero vector of $V_{\Omega}$ has such an element 
of the loop algebra 
that maps it to $\Omega$. Suppose that there is a nonzero vector $v_n$ 
in the sector $h_0= r-2n$ of $V_{\Omega}$ that has no such element. 
Then, we have  
 \be 
 x_{k_1}^{+}\cdots x_{k_n}^{+} \, v_n =0
\label{eq:vn0}
 \ee 
for all monomial elements $x_{k_1}^{+}\cdots x_{k_n}^{+}$.  
Let us express $v_n$ in terms of the basis vectors 
$\rho_{j_1}({\bm a}) \cdots \rho_{j_n}({\bm a}) \Omega$ with 
coefficients $C_{j_1, \ldots, j_n}$ 
such as shown in eq. \ref{eq:vnC}. Then,  
by the same argument as in lemma \ref{lem:vnC} 
using the Vandermonde determinant, we show from eq. (\ref{eq:vn0}) that 
all the coefficients $C_{j_1, \ldots, j_n}$ vanish. 
However, this contradicts with the assumption that 
$v_n$ is nonzero. It therefore follows that $v_n$ has such an element 
that maps it to $\Omega$. We thus obtain the theorem.  
\end{proof}

\subsection{The case of degenerate evaluation parameters}

Let us discuss a general criteria for a finite-dimensional 
highest weight representation to be irrducible. 
\begin{th1} 
Let $\Omega$ be a highest weight vector that has 
evaluation parameters $a_j$ with multiplicities 
$m_j$ for $j=1, 2, \ldots, s$.  
We denote by ${\bm a}$ the set of evaluation parameters, i.e.  
${\bm a}=\{a_1, a_2, \ldots, a_s  \}$, and by $V_{\Omega}$  
 the representation  generated by $\Omega$. Then,  
$V_{\Omega}$  is irreducible if and only if 
$x_{s}^{-}({\bm a}) \Omega = 0$.   
\label{th:degenerate}
\end{th1}
We prove it by generalizing 
the proof of theorem \ref{th:distinct}. However, 
we shall discuss the derivation in the next report \cite{loop}.

Let us discuss an application of theorem \ref{th:degenerate} 
for the degenerate case. 
It plays an important role when we consider 
the spectral degeneracy of the XXZ spin chain at roots of unity 
associated with the $sl_2$ loop algebra symmetry.  
It has been shown that a regular Bethe ansatz eigenvector $|R \ra$ 
in the sector (\ref{sct0}) is a highest weight vector \cite{HWT}. 
However, if the evaluation parameters 
of the regular Bethe state $|R \ra$ 
are not distinct, it is nontrivial whether the 
highest weight representation $V_{|R \ra}$ is irreducible or not. 
Suppose that we have the following relation for $|R \ra$:  
\be 
x_{s}^{-}({\bm a}) \, |R \ra = 0 \, , 
\label{eq:criteria}
\ee 
where ${\bm a}$ denotes 
the set of evaluation parameters $a_1, a_2, \ldots, a_s$ 
of the highest weight vector $|R \ra$.  
Then, it follows from theorem \ref{th:degenerate} that 
the highest weight representation $V_{|R \ra}$ is irreducible.

In the most general cases, however, 
it is not clear whether the condition (\ref{eq:criteria}) 
holds for all the regular Bethe eigenstates in the sector (\ref{sct0}).  
In order to discuss the degenerate multiplicity of the $sl_2$ 
loop algebra symmetry of the XXZ spin chain, 
 it is thus important to know the dimensions of reducible 
finite-dimensional highest weight representations. We propose the 
following conjecture: 
\begin{conj}
 Let  $\Omega$ be a highest weight vector with 
 evaluation parameters 
 ${\hat a}_1, {\hat a}_2, \ldots, {\hat a}_r$.  
If $x^{-}_{r-1}({\hat A}_j) \Omega \ne 0$  for all $j$, 
the representation $V_{\Omega}$ generated by  $\Omega$ 
is reducible and indecomposable, and it has the dimensions $2^r$.
Here ${\hat A}_j$ denote the set 
${\hat A}_{\phi} \setminus \{ {\hat a}_j \}$ 
with ${\hat A}_{\phi}= \{ {\hat a}_j | j=1, 2, \ldots, r \}$. 
\end{conj}

For an illustration of reducible indecomposable 
highest weight representations, 
let us consider the case of $r=3$ with $m_1=2$ and $m_2=1$. 
In this case,  the highest weight vector $\Omega$ has evaluation parameters, 
${\hat a}_1=a_1$, ${\hat a}_2=a_1$ and ${\hat a}_3=a_2$, i.e. 
${\hat A}_{\phi}=\{ a_1, a_1, a_2 \}$.  
Here we recall lemma \ref{lem:sector}. The highest weight 
representation $V_{\Omega}$ has four sectors of 
$h_0=3, 1, -1, 3$, respectively. 
Let us introduce some symbols. 
\be 
\rho_1 = x_1^{-}(a_2) \, , \quad 
\rho_2 = x_1^{-}(a_1) \, , \quad 
w_1 = x_2^{-}(a_1, a_2) \, . 
\ee
Here $x_2^{-}(a_1, a_2)$ denotes 
 $x_2^{-}(B)$ with $B=\{a_1, a_2 \}$. 
It is easy to show from the reduction relation,  
$x_3^{-}({\hat A}_{\phi})\Omega = 0$,  that 
$x_n^{-} \Omega$ is expressed in terms of $\rho_1 \Omega$,  
$\rho_2 \Omega$, and $w_1 \Omega$ as follows. 
\be
x_n^{-} \Omega = {\frac {a_1^n} {a_{12}}} \, \rho_1 \Omega +  
{\frac {a_2^n} {a_{21}}} \, \rho_2 \Omega +
\left( {\frac {n a_1^{n-1}} {a_{12}}}
- {\frac {a_1^n -a_2^n} {a_{12}}} \right) \,  w_1 \Omega 
\ee
We now have  
\be 
x_n^{+} w_1 \Omega = 0 \, , \quad {\rm for} \quad n \in {\bf Z}  \, . 
\ee 
Therefore, it follows that the representation $V_{\Omega}$ 
is reducible and indecomposable if $w_1 \Omega \ne 0$.   
By a similar argument with lemma \ref{lem:square0} we show  
\be \rho_1^3 \Omega \ne 0 \, , \quad 
\rho_1^4 \Omega =0;  \quad 
\rho_2 \Omega \ne 0\, , \quad \rho_2^2 \Omega =0;   
\quad w_1^2 \Omega = 0. 
\ee
We also show 
\be 
\rho_1 w_1 \Omega= 0 \, ,     
\ee
and $\rho_2 w_1 \Omega \ne 0$ if $w_1 \Omega \ne 0$.  
It thus follows that  the four sectors of $V_{\Omega}$ have  
 the following sets of basis vectors: 
\be
\begin{array}{cccc} 
  \Omega , &  &  & {\rm for} \quad h_0=3 ; \non \\ 
 \rho_1 \Omega , & \rho_2 \Omega , &  w_1 \Omega , & 
\quad  {\rm for} \quad h_0=1 ; \non \\ 
  \rho_1^2 \Omega, & \rho_1 \rho_2 \Omega, & \rho_2 w_1 \Omega , & 
\quad  {\rm for} \quad h_0=-1 ; \non \\    
 \rho_1^3 \Omega , & & & \quad  {\rm for} \quad h_0=-3 \, .   
\end{array}
\ee
Consequently, we have the following result: 
\begin{prop} 
The highest weight representation with three evaluation parameters 
$a_1$, $a_1$,  $a_2$ is reducible, indecomposable 
and of $2^3$ dimensions, if and only if $w_1 \Omega \ne 0$. 
It is irreducible and of 6 dimensions,  
if and only if $w_1 \Omega = 0$. 
\end{prop}

\subsection*{Acknowledgements}

The author would like to thank 
the organizers for their kind invitation to the international workshop 
``Recent Advances in Quantum Integrable Systems'', September 6-9, 
2005, LAPTH, Annecy-le-Vieux, France.  
He would also like to thank many participants of the conference for 
useful comments and discussions. 
This work is partially supported by the Grant-in-Aid (No. 17540351).

\end{document}